\documentclass{epl}

\title{On the role of hydrodynamic interactions in colloidal gelation}
\author{R. Yamamoto\inst{1,2} \and K. Kim\inst{1,2} \and Y. Nakayama\inst{3}
\and K. Miyazaki\inst{4} \and D.R. Reichman\inst{4}}
\institute{
  \inst{1} Department of Chemical Engineering, Kyoto University, Kyoto 615-8510, Japan\\
  \inst{2} PRESTO, Japan Science and Technology Agency, Kawaguchi 332-0012, Japan\\ 
  \inst{3} Department of Chemical Engineering, Kyushu University, Fukuoka 819-0395, Japan\\
  \inst{4} Department of Chemistry, Columbia University, 3000 Broadway,
New York NY 10027, USA}
\pacs{47.57.-s}{Complex fluids and colloidal systems}
\pacs{82.20.Wt}{Computational modeling; simulation}
\pacs{83.10.Pp}{Particle dynamics} 
\pacs{83.80.Kn}{Physical gels and microgels}
\begin{document}

\maketitle

\begin{abstract}
In this letter, we investigate several aspects related to the effect of
 hydrodynamics interactions on phase separation-induced gelation of
 colloidal particles. 
We explain physically the observation of Tanaka and Araki
[Phys. Rev. Lett. {\bf 85}, 1338 (2000)] of hydrodynamic stabilization of
 cellular network structures in two dimensions. 
We demonstrate that hydrodynamic interactions have only a minor
 quantitative influence on the structure of transient gels in three
 dimensions. 
We discuss some experimental implications of our results. 
\end{abstract}

There has been a surge of interest in recent years in understanding
the dynamics and rheological properties of colloidal
gels\cite{Chen1991,Rueb1997,Bergenholtz1999}.   
Such systems may be formed under a wide variety of conditions, ranging
from percolation\cite{miller2003,deCandia2005b},  
nearly arrested phase
separation\cite{Verhaegh1997,GeoffreySoga1998,Foffi2005,Manley2005}, a
glass-like transition 
involving extended structures\cite{Kroy2004,Sciortino2004}, and
diffusion-limited cluster aggregation\cite{Carpineti1992}.  
The structure and rheology of the resultant gel depends
sensitively on the volume fraction, the region of the phase diagram in
which the system is initially prepared, and the range and strength of
the interactions between the colloidal particles.  Modern simulation
techniques can realistically capture many of the features that are
observed experimentally.

One aspect of the theoretical modeling of colloidal gelation that is
incomplete is our understanding of the role of hydrodynamic
interactions \cite{happel1983,kim1991,russel1989} 
in determining the gel morphologies that may form under a
variety of conditions.  Hydrodynamic interactions are intrinsically
many-body in nature, and simple low-order multipole approximations are
not sufficiently accurate to yield a reliable account of the effect of
solvent mediated forces on colloidal particles as gelation proceeds.
Several simulation approaches have been developed to properly account
for the many-body nature of hydrodynamic interactions.  The
Accelerated Stokesian Dynamics (ASD) method \cite{sierou2001}
of Brady and coworkers
uses a multipole-like decomposition and separation of near-field and
far-field components to achieve accuracy and computational efficiency.
While this method scales better with system size than older Stokesian
approaches \cite{brady1988}, it is still difficult to use 
this approach for Brownian
systems involving more than 500-1000 particles.  Another useful
approach is the Lattice Boltzmann (LB) method \cite{Ladd2001,Cates2005c}.  
LB is a grid-based
method that operates at the level of the reduced phase space
distribution function.  The LB approach is capable of handling large
system sizes efficiently.  One difficultly with LB is that the
coarse-grained parameters that enter the simulation must be determined
with care, and, in general, artifacts are sometimes difficult to
detect\cite{Ladd2001,Cates2005c}.  
While the LB the method must be used with care for
multi-component systems, impressive results have been recently been
obtained\cite{Stratford2005}.

In this work we will take another approach, based on direct solution
of the Navier-Stokes equation on a grid.  One difficulty with such a
direct approach is the treatment of the interface between the fluid
and the colloidal particle.  If the colloidal particle is treated as a
non-deformable solid, the need to adapt the grid at this interface as
the colloid position is changed makes this approach expensive
computationally.  Recently, several approaches have been developed to
overcome this difficulty. 
Tanaka  and Araki have developed an approach
where the colloidal particle is treated as a high viscosity fluid
\cite{Tanaka2000b}. 
This obviates the grid adjustment issue, but at the expense of
drastically decreasing the time step for evolving the colloidal
particle's position.  More recently, Yamamoto and coworkers have
developed a smooth-profile interface approach\cite{Nakayama2005}, which
also does not require grid adjustment.  
Furthermore, this approach does not require small time steps for the
evolution of the particle position. 
This approach make it possible to simulate the evolution of large systems
of colloidal particles interacting with full many-body hydrodynamic
interactions for reasonably long time scales.  This is the approach
taken here.

In the work of Tanaka  and Araki\cite{Tanaka2000b}, 
a preliminary investigation of the
role of hydrodynamics in the formation of transient gels was made, and
it was concluded that hydrodynamic interactions serve to stabilize
tenuous spanning networks and structures.  
All reported studies in the work of Tanaka  and Araki were 
performed in two dimensions. 
In this work,  
we study larger system sizes, including three dimensional 
systems composed of more than 2000 colloidal particles, 
using the method of Refs.\cite{Nakayama2005,Kim2006,Nakayama2006}.  
Our goals are twofold: first we
provide a physical picture for the origin of the effect observed by
Tanaka  and Araki.  
Secondly, we investigate how robust this effect is
with respect to dimensionality.  We conclude this work with a brief
discussion of the observability of the effects discussed here in actual
experiments.

We study systems of volume fraction $\phi =0.173$ and $\phi=0.338$ in
two dimensions and $\phi=0.153$ and $\phi=0.307$ in three dimensions.
Two dimensional systems are simulated on a 512$\times$512 grid and contain
400 and 784 particles, respectively.  Three dimensional systems are
performed on a 128$\times$128$\times$128 grid and contain 1200 and 2400 particles,
respectively.  The colloidal suspension is mono-disperse, and colloidal
particles interact via a standard Lennard-Jones potential
$V(r)=4\epsilon\{ (\sigma/r)^{12}-(\sigma/r)^{6}\}$. 
All
systems are equilibrated at high temperature, and then instantaneously
quenched to zero temperature, thus Brownian motion is ignored.  This
quench protocol may be relaxed so that Brownian motion is included
within the approach of Refs.\cite{Nakayama2005,Kim2006,Nakayama2006},
but the qualitative features that will be discussed in this work are
unaltered if finite temperature quenches are performed.  The numerical
approach for resolving hydrodynamic interactions between colloidal
particles is detailed fully in
Refs.\cite{Nakayama2005,Kim2006,Nakayama2006} and will not be discussed
here. 

Fig. 1 shows evolution of particle aggregation after a quench for the
low volume fraction two dimensional system.  The left column shows two
snapshots of this evolution without hydrodynamic interactions.  The
right column shows the same stages of evolution for the same system
with hydrodynamics.  Solvent colour provides a representation of the
pressure of the fluid, with red denoting highest pressure, and blue
lowest pressure.  Once the pressure becomes uniform across the sample,
the evolution of the cluster morphologies are rather similar with and
without hydrodynamic interactions, at least at intermediate times.
This is illustrated in Fig.2, where the average number of bonds per
particle as a function of time during the evolution of the aggregation
process.
Here the bond is defined as a pair of 
particles whose distance from each
other is less than $1.2\sigma$.

For higher volume fractions in two dimensions, a qualitative
difference may be observed.  Fig.3 shows a similar comparison of the
evolution of domains in a colloidal system quenched to $T=0$ at the
volume fraction $\phi=0.338$.  Evolution without hydrodynamics leads
eventually to disjoint clusters composed of relatively compact
structures.  Upon including the effect of hydrodynamic interactions,
long-lived cellular structures may be observed.  The cells of the gel
morphology contain fluid of uniform pressure, but the fluid pressure
is inhomogeneous globally, even for times where the pressure has
completely uniformized at the volume fraction $\phi=0.173$.  This
pressure (force) balance between colloidal particles that make up the
walls of cellular structures and the fluid contained within them leads
to a noticeable slowing down of the evolution of coarsening and a
concomitant stabilization of tenuous gel structure.  This is clearly
illustrated in Fig.4, where the time-dependent structure factor
$S(k,t)$ is calculated with and without hydrodynamics at various times
after the initial quench.  Observation of the low wave vector peak in
$S(k,t)$ demonstrates clearly that the inclusion of hydrodynamic
interactions stabilize gels composed of cellular structures.  At longer
times, such structure will eventually break up.  While this break-up
will be accelerated by Brownian motion, it may also be retarded by
inducing, via a depletion interaction, a short-ranged strong
attraction between the colloidal particles.  Indeed, recent work has
demonstrated that even in the absence of hydrodynamics, gels may be
formed as via a glass-like arrest of the colloid dense phase a result
of quenching systems with very strong, short-range
attractions\cite{Verhaegh1997,GeoffreySoga1998,Foffi2005,Manley2005}. 

We now turn to the three dimensional case.  Fig.5 shows both low (left
column, $\phi=0.153$) and higher (right column, $\phi=0.307$) volume
fractions.  The upper panel in each column shows evolution without
hydrodynamics while the lower panel in each column corresponds to
evolution with hydrodynamics.  In three dimensions it is clearly
difficult to detect gross differences in particle configurations that
evolve with and without hydrodynamics.  Fig.6 shows the average number
of bonds per particle in both the low and higher volume fraction
cases.  Similar to the two dimensional case, at short to intermediate
times hydrodynamics leads to slower growth and slightly more tenuous
structure.  A qualitative difference between two and three dimensions,
however, may be seen by examining the evolution of $S(k,t)$ (not shown
in 3D).  
While it is clear that the evolution is different with and without
hydrodynamic interactions in three dimensions, no time interval is found
where the 
structure ceases to evolve, even transiently.

While the range of hydrodynamic interactions is drastically different
in two and three dimensions, the origin of the network-stabilizing
effect in two dimensions appears to be geometric.  In two dimensions
cellular structures may form naturally during phase separation, and
then become transiently locked in via the pressure balance
effect discussed above.  In three dimensions, the typical situation
will involve bicontinuity of colloidal structure and the suspending
fluid which allows for global fluid pressure equilibration.  Thus,
hydrodynamics is expected to play a less important role in three
dimensions.  This reasoning suggest that even in the realistic,
quasi-two dimensional case, where momentum may be dissipated in the
third dimension, the network stabilizing effect may exist in some form
even though the effective range of interactions is actually {\em shorter} in
the quasi-two dimensional case than it is in three
dimensions\footnote{Free standing films might better approximate the
pure two dimensional case. One possibility would involve bubble dynamics
in such a film\cite{Diamant_2006private}}\cite{Diamant2005}. 

In conclusion, the role of hydrodynamics in the formation of colloidal
gels has been studied.  Using the smoothed profile
approach\cite{Nakayama2005,Kim2006,Nakayama2006}, large system sizes in
two and 
three dimensions have been studied.  A physical explanation for the
transient arrest of gel structure in two dimensions as noted by 
Tanaka  and Araki has been given.  The effect of hydrodynamics in
influencing gel structure is found to be significantly greater in two
dimensions, where cellular structures form naturally.  The observation
that cellular structures may transiently lead to inhomogeneous
pressure profiles in the suspending fluid that drastically slow the
evolution of gel morphology suggests that such an effect may be seen
in confined geometries on some time scale, even if the range of
interaction in the 
quasi-two dimensional case is significantly shorter than in the pure
two and three dimensional cases.

\acknowledgements

RY thank the Earth Simulator Center, Japan Agency for Marine-Earth
Science and Technology (JAMSTEC) for their generous supports to this
work. 
DRR would like to thank M. E. Cates, H. Diamant, and E. Rabani for
useful discussions. 
DRR and KM have been supported by NSF (\# 0134969).

\newpage

\begin{figure}
\oneimage[width=130mm]{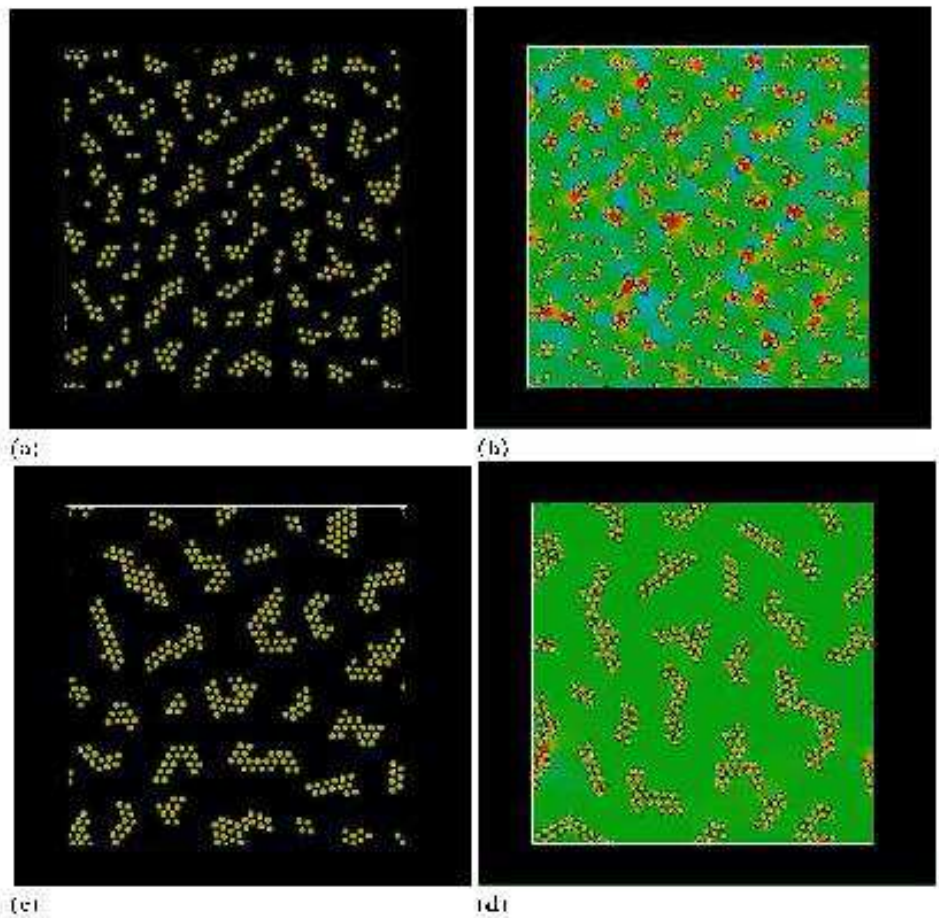}
\caption{Snapshots of aggregating particles in a colloidal dispersion
 without [early time (a) and late time (c)] or with [early time (b) and
 late time (d)] hydrodynamic interactions. Here the volume fraction of 
 particles is 0.173 ($N=400$). The background colour in (b) and (d)
 represents the pressure of the fluid. Blue and red correspond to low
 and high pressure, respectively.} 
\label{f.1}
\end{figure}

\begin{figure}
\oneimage[width=65mm]{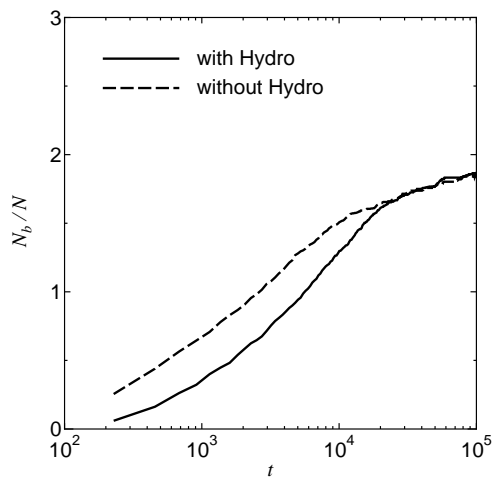}
\caption{Time dependence of the average number of bonds per 
a particle in two dimensions for $\phi=0.173$}
\label{f.2}
\end{figure}

\begin{figure}
\oneimage[width=130mm]{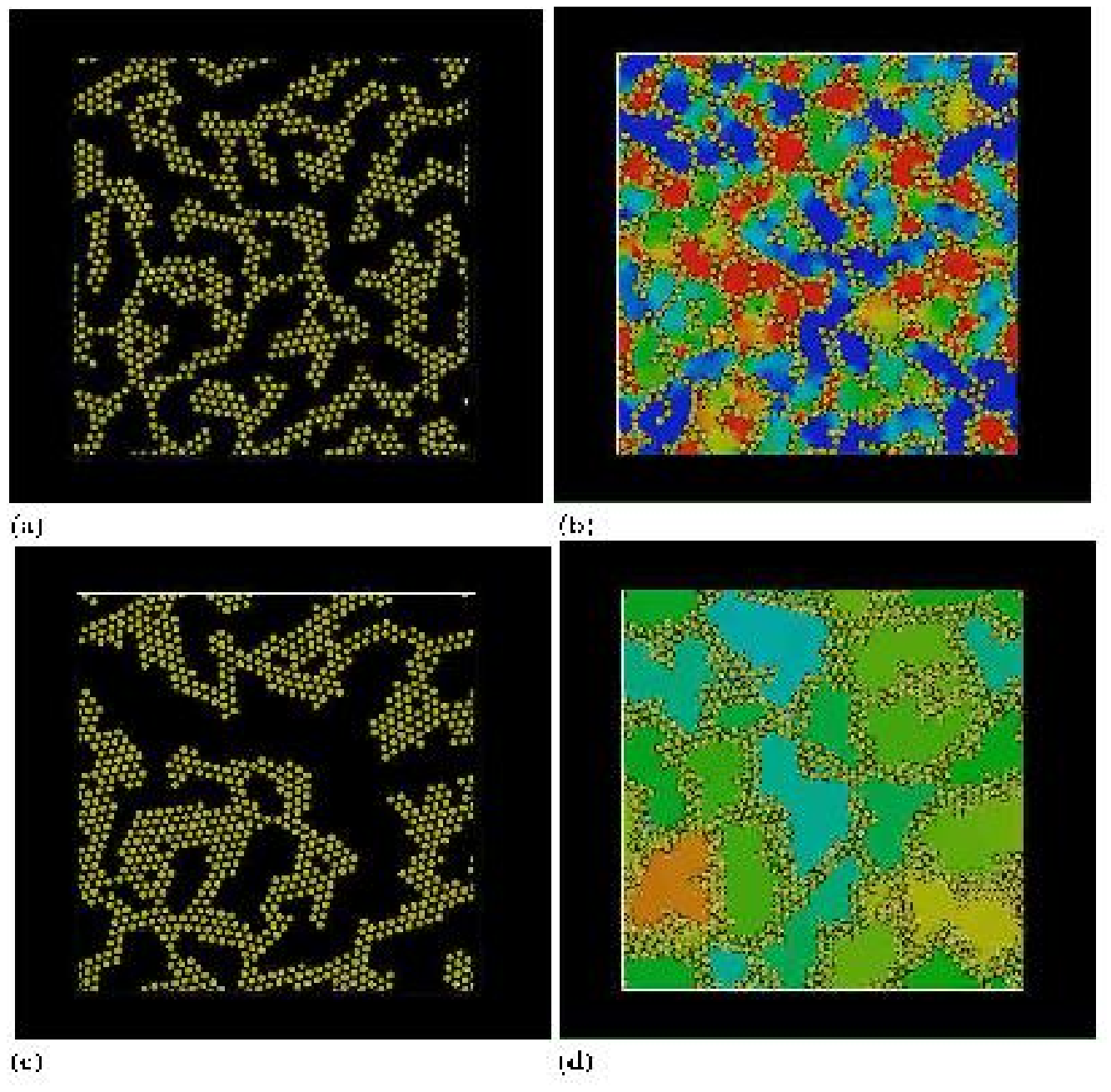}
\caption{Snapshots of aggregating particles in a colloidal dispersion
 without [early time (a) and late time (c)] or with [early time (b) and
 late time (d)] hydrodynamic interactions. 
Here the volume fraction of particles is 0.338 ($N=784$). The background
 colour in (b) and (d) represents the pressure of the fluid. Blue and red
 correspond to low and high pressure, respectively.} 
\label{f.3}
\end{figure}

\begin{figure}
\twoimages[width=65mm]{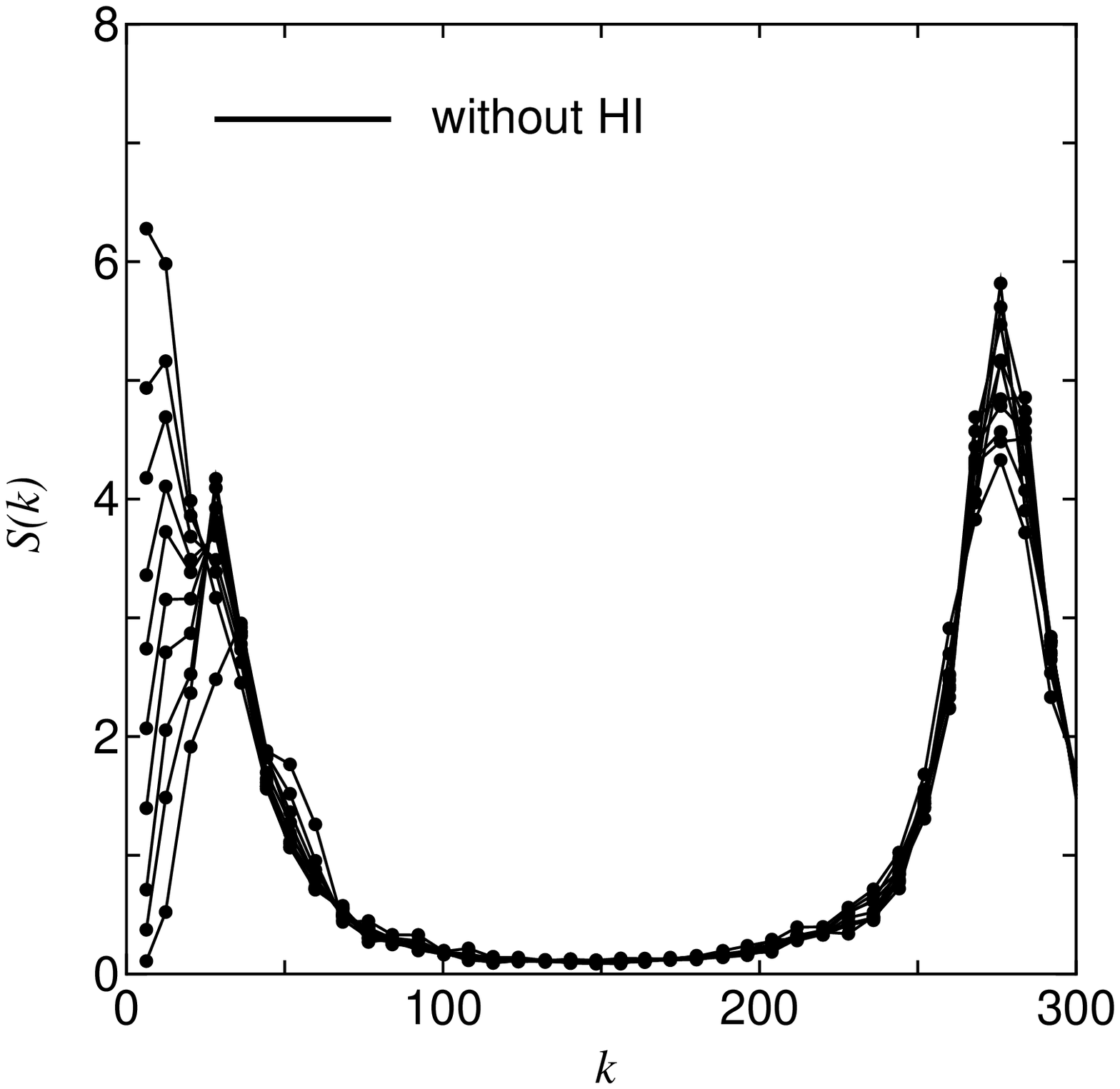}{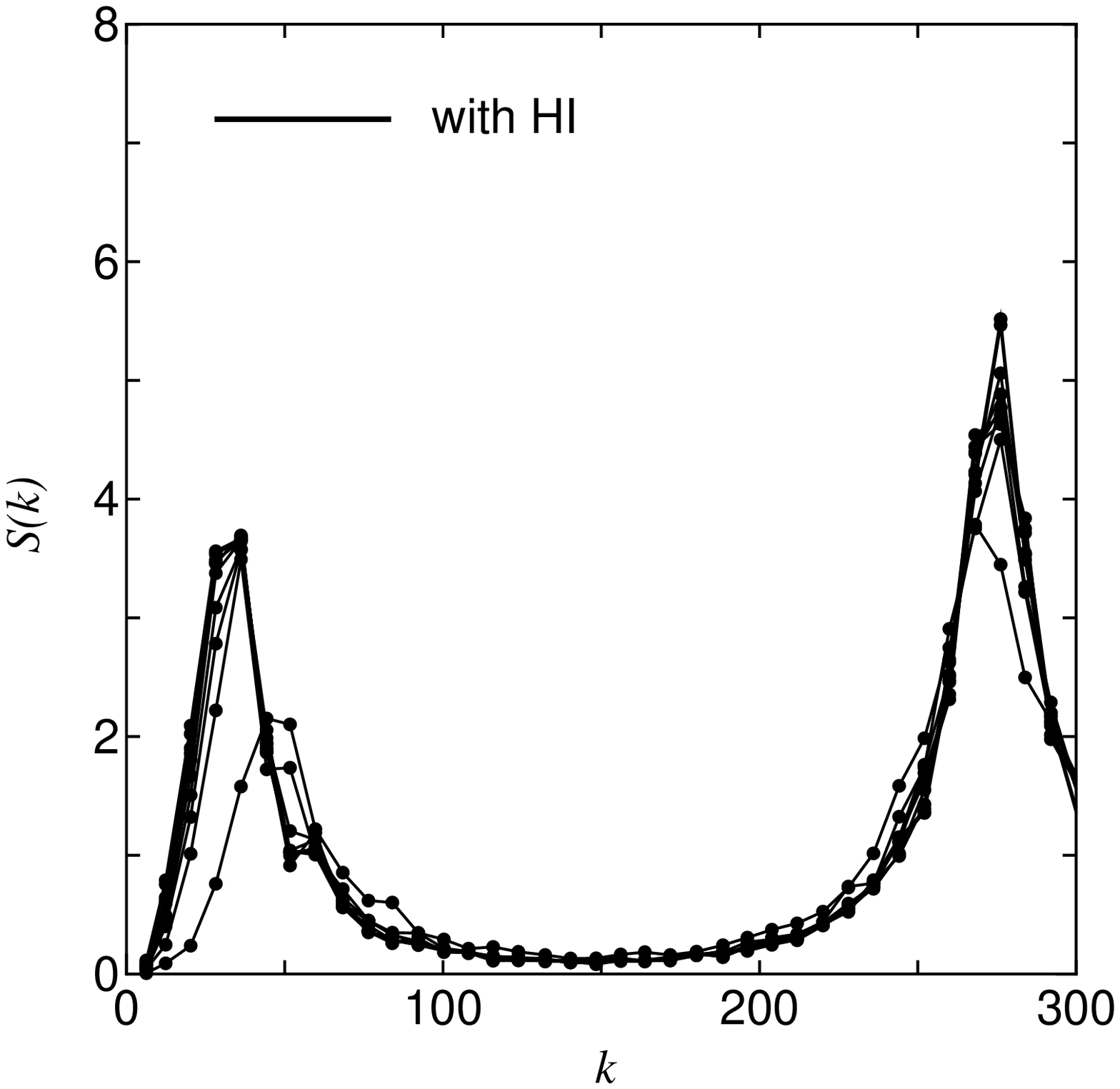}
(a)\hspace{67mm}(b)
\caption{S(k) without (a) and with (b) hydrodynamic interactions for
 $\phi=0.338$.
10 lines are plotted from $t=1.5\times 10^3$ (lowest) to $1.5\times 10^4$
 (highest) for every $\delta t=1.5\times 10^3$. 
Here the time unit is 
is $\rho \Delta2 / \eta$, where $\rho$ is the fluid mass density, $\eta$
 is the fluid viscosity, and $\Delta$ is the grid spacing (=1). 
}
\label{f.4}
\end{figure}

\begin{figure}
\oneimage[width=130mm]{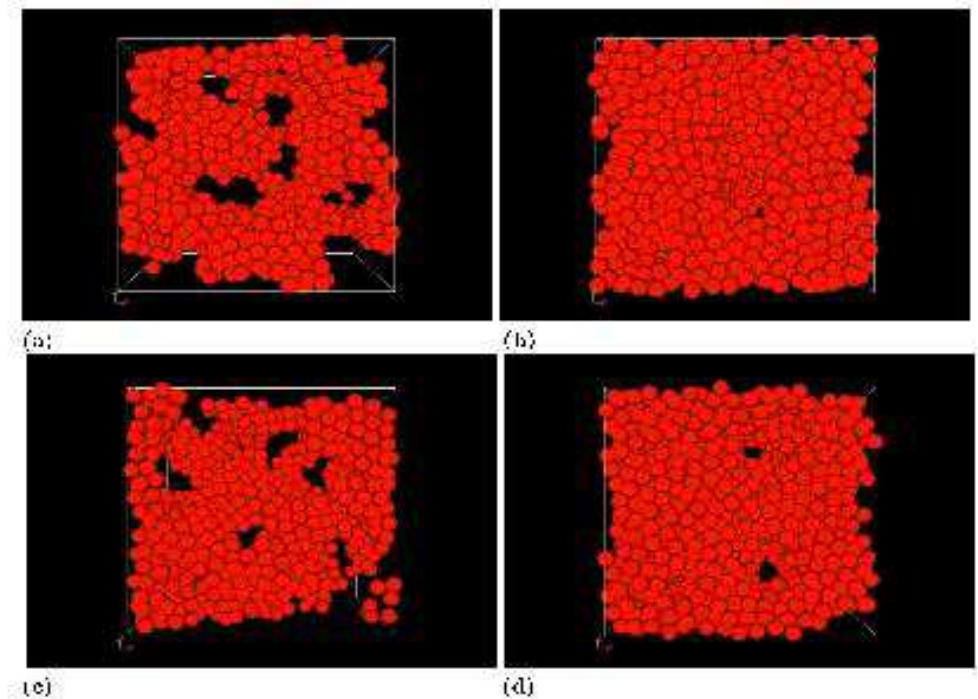}
\caption{Snapshots of aggregating particles in a three dimensional
 dispersion without 
 [$\phi=0.153$ (a) and $\phi=0.307$ (b)] or 
with [$\phi=0.153$ (c) and  $\phi=0.307$ (d)] 
hydrodynamic interactions.
All configurations are obtained at $t = 1.5 \times 10^4$.
}
\label{f.5}
\end{figure}

\begin{figure}
\twoimages[width=65mm]{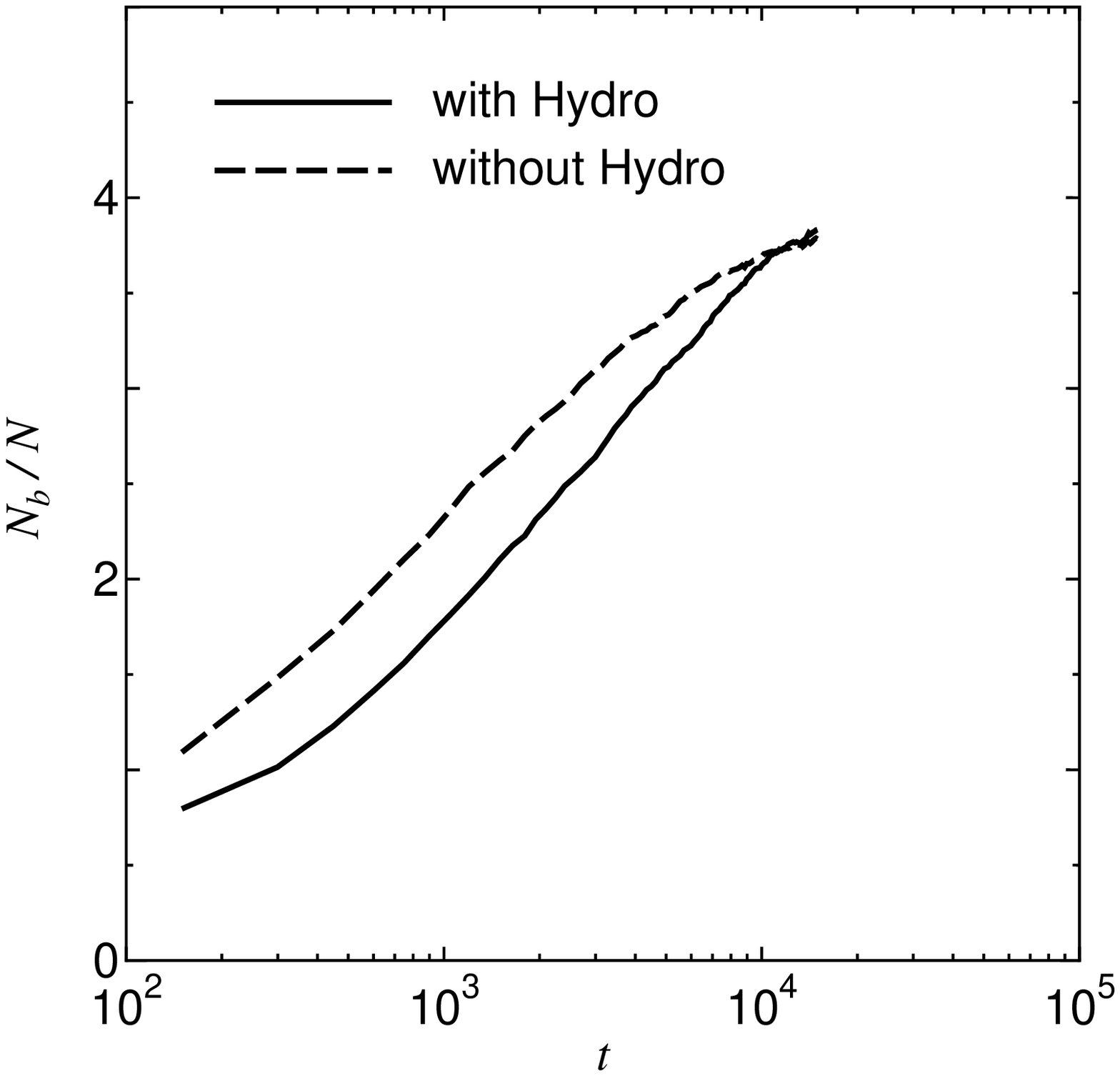}{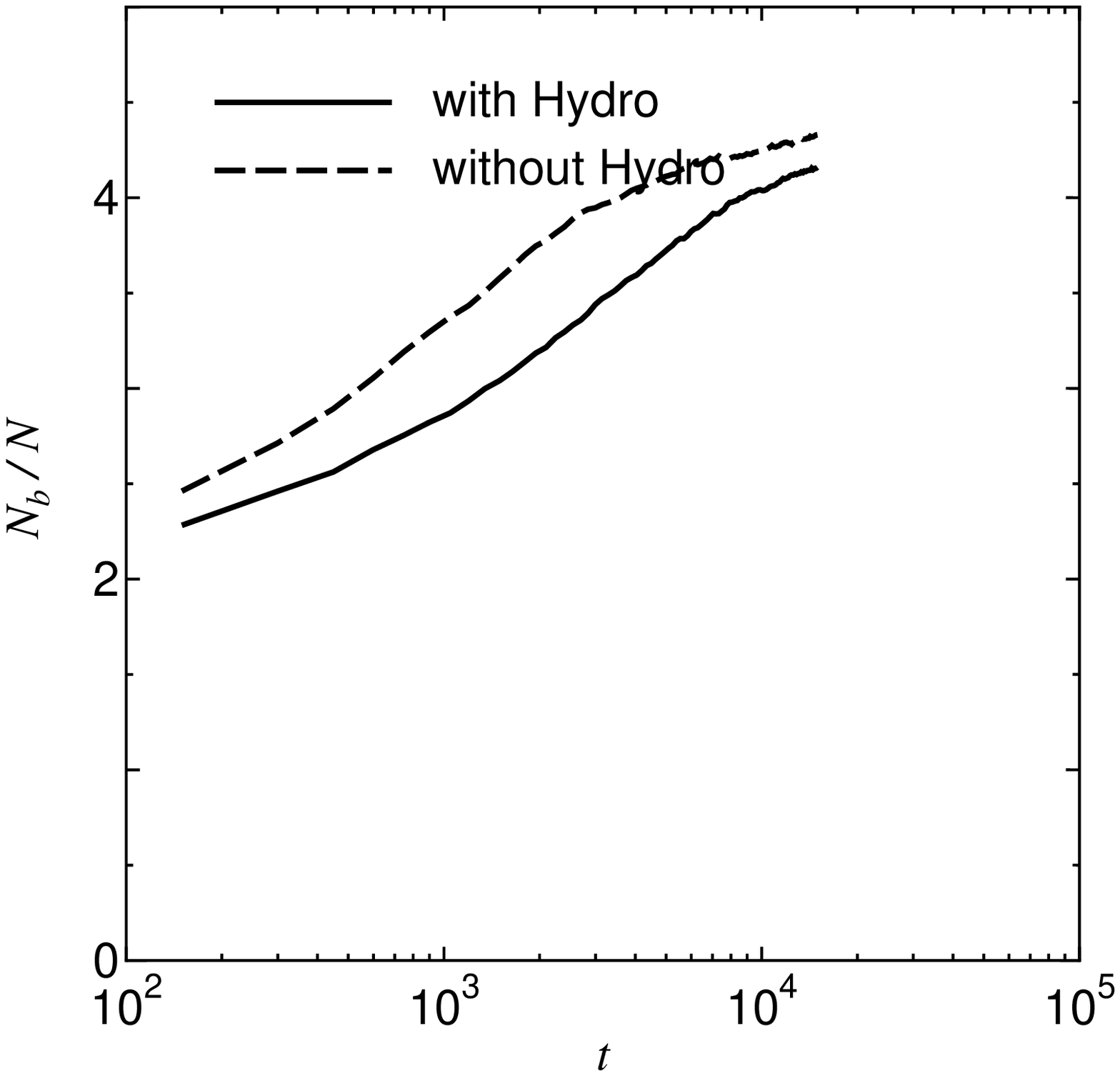}
(a)\hspace{67mm}(b)
\caption{Time dependence of the average number of bonds per a particle
 in three dimensions for  $\phi=0.153$ (a) and $0.338$ (b).}
\label{f.6}
\end{figure}


\begin{thebibliography}{10}

\bibitem{Chen1991}
\Name{Chen M.~ \and Russel W.~B.}
\REVIEW{J. Colloid Interface Sci.}{141}{1991}{564}.

\bibitem{Rueb1997}
\Name{Rueb C.~J. \and Zukoski C.~F.}
\REVIEW{J. Rheol.}{41}{1997}{197}.

\bibitem{Bergenholtz1999}
\Name{Bergenholtz J. \and Fuchs M.}
\REVIEW{Phys. Rev. {\rm E}}{59}{1999}{5706}.

\bibitem{miller2003}
\Name{Miller M.~A. \and Frenkel D.}
\REVIEW{Phys. Rev. Lett.}{90}{2003}{135702}.

\bibitem{deCandia2005b}
\Name{{de Candia} A., {Del Gado} E., Fierro A., Sator N. \and Coniglio A.}
\REVIEW{Physica {\rm A}}{358}{2005}{239}.

\bibitem{Verhaegh1997}
\Name{Verhaegh N.~A.~M., Asnaghi D., Lekkerkerker H.~N.~W., Giglio
 M. \and Cipelletti, L.}
\REVIEW{Physica {\rm A}}{242}{1997}{104}.

\bibitem{GeoffreySoga1998}
\Name{{Geoffrey Soga} K., Melrose J.~R. \and Ball R.~C.}
\REVIEW{\em J. Chem. Phys.}{108}{1998}{6026}.

\bibitem{Foffi2005}
\Name{Foffi G., {De Michele} C.,  Sciortino F. \and Tartaglia P.}
\REVIEW{Phys. Rev. Lett.}{94}{2005}{078301}.

\bibitem{Manley2005}
\Name{Manley S.,  Wyss H.~M., Miyazaki K.,  Conrad J.~C. Trappe V.,
 Kaufman L.~J., Reichman D.~R. \and Weitz D.~A.}
\REVIEW{Phys. Rev. Lett.}{95}{2005}{238302}.

\bibitem{Kroy2004}
\Name{Kroy K.,  Cates M.~E. \and Poon W.~C.~K.}
\REVIEW{Phys. Rev. Lett.}{92}{2004}{148302}.

\bibitem{Sciortino2004}
\Name{Sciortino F., Mossa S., Zaccarelli E. \and Tartaglia P.}
\REVIEW{Phys. Rev. Lett.}{93}{2004}{055701}.

\bibitem{Carpineti1992}
\Name{Carpineti M. \and Giglio M.}
\REVIEW{Phys. Rev. Lett.}{68}{1992}{3327}.

\bibitem{happel1983}
\Name{Happel J. \and Brenner H.}
\Book{Low Reynolds number hydrodynamics: with special applications
  to particulate media}.
\Publ{Kluwer Academic Publishers, Dordrecht}
\Year{1983}.

\bibitem{kim1991}
\Name{Kim S. \and Karrila S.~J.}
\Book{Microhydrodynamics: Principles and Selected Applications}.
\Publ{Butterworth-Hwinemann, Boston}
\Year{1991}.

\bibitem{russel1989}
\Name{Russel W.~B., Saville D.~A. \and Schowalter W.~R.}
\Book{Colloidal Dispersions}.
\Publ{Cambridge University Press, Cambridge}
\Year{1989}.

\bibitem{sierou2001}
\Name{Sierou A. \and Brady J.~F.}
\REVIEW{J. Fluid Mech.}{448}{2001}{115}.

\bibitem{brady1988}
\Name{Brady J.~F. \and Bossis G.}
\REVIEW{Ann. Rev. Fluid Mech.}{20}{1988}{111}.

\bibitem{Ladd2001}
\Name{Ladd A.~J.~C. \and Verberg R.}
\REVIEW{J. Stat. Phys.}{104}{2001}{1191}.

\bibitem{Cates2005c}
\Name{Cates M.~E., Desplat J.-C., Stansell P., Wagner A.~J., Stratford
 K., Adhikari R. \and Pagonabarraga I.}
\REVIEW{Phil. Trans. R. Soc. {\rm A}}{363}{2005}{1917}.

\bibitem{Stratford2005}
\Name{Stratford K., Adhikari R., Pagonabarraga I., Desplat J.-C. \and
 Cates M.~E.}
\REVIEW{Science}{309}{2005}{2198}.

\bibitem{Tanaka2000b}
\Name{Tanaka H. \and Araki T.}
\REVIEW{Phys. Rev. Lett.}{85}{2000}{1338}.

\bibitem{Nakayama2005}
\Name{Nakayama Y. \and Yamamoto R.}
\REVIEW{Phys. Rev. {\rm E}}{71}{2005}{036707}.

\bibitem{Kim2006}
\Name{Kim K., Nakayama Y. \and Yamamoto R.}
\REVIEW{\tt cond--mat/0601534}{}{2006}{}.

\bibitem{Nakayama2006}
\Name{Y.~Nakayama, K.~Kim, and R.~Yamamoto}
\REVIEW{\tt cond-mat/0601322}{}{2006}{}.

\bibitem{Diamant2005}
\Name{Diamant H., Cui B., Lin B. \and Rice S.~A.}
\REVIEW{J. Phys.: Condens. Matter}{17}{2005}{S2787}.

\bibitem{Diamant_2006private}
\Name{Diamant H.}
\Book{{\rm (private communication)}}.

\end{thebibliography}
\end{document}